\documentclass[twoside]{article}
\usepackage[dvips]{graphicx}
\usepackage{fleqn,espcrc2}

\usepackage[figuresright]{rotating}

\newcommand{\be}{\begin{equation}}
\newcommand{\ee}{\end{equation}}
\newcommand{\bc}{\begin{center}}
\newcommand{\ec}{\end{center}}
\newcommand{\bes}{\begin{equation*}}
\newcommand{\ees}{\end{equation*}}
\newcommand{\beqn}{\begin{eqnarray}}
\newcommand{\eeqn}{\end{eqnarray}}
\newcommand{\beqns}{\begin{eqnarray*}}
\newcommand{\eeqns}{\end{eqnarray*}}
\newcommand{\bs}{B_s^0\!-\!\bar B_s^0}
\newcommand{\widdif}{\frac{\Delta \Gamma_{B_s}}{\Gamma_{B_s}}}
\newcommand{\barms}{\overline{\mathrm{MS}}}
\newcommand{\err}[2]{\raisebox{0.08em}{\scriptsize {$\;\begin{array}{@{}l@{}}
			  \plus\makebox[2.05em][r]{#1} \\[-0.12em] 
			  \minus\makebox[2.05em][r]{#2} 
			\end{array}$}}}

\newcommand{\errd}[2]{\raisebox{0.08em}{\scriptsize {$\;\begin{array}{@{}l@{}}
			  \plus\makebox[1.35em][r]{#1} \\[-0.112em] 
			  \minus\makebox[1.35em][r]{#2} 
			\end{array}$}}}
\newcommand{\plus}{\makebox[2pt][c]{$+$}}
\newcommand{\minus}{\makebox[2pt][c]{$-$}}

\newcommand{\AmS}{{\protect\the\textfont2
  A\kern-.1667em\lower.5ex\hbox{M}\kern-.125emS}}

\title{Width difference in the $\bs$ system from lattice HQET}

\author{V. Gim\'enez\address{Dep. de F\'\i sica Te\'orica, IFIC and Univ. de
Valencia,\\  
Dr. Moliner 50, E-46100, Burjassot, Valencia, Spain} and J. Reyes\addressmark
\thanks{Govierno Vasco fellowship.}\thanks{Talk presented by J. Reyes.}}
\begin{document}

\begin{abstract}
We present recent results for the prediction of the $\bs$ lifetime difference
from lattice Heavy Quark Effective Theory simulations. In order to get a
next-to-leading order result  we have calculated the matching between QCD
and HQET and the two loop anomalous dimension in the HQET for all the $\Delta B=2$
operators, in particular for the operators which enter in the width difference.
We obtain for the $\bs$ lifetime difference, $\widdif=(5.1\pm 1.9\pm 1.7)\times
10^{-2}$.
\vspace{1pc}
\end{abstract}

\maketitle

\section{Introduction.}
The width difference ($\widdif$) in the $\bs$ system is expected to be the
largest among bottom hadrons. The Standard Model prediction for $\widdif$
relies on an operator product expansion, where the short distance 
scale is provided by the $b$ quark mass \cite{ben97}. The theoretical
 estimations for $\widdif$ are in the range $5\div 15 \%$ and hadronic matrix
elements of four quark operators are crucial inputs. We present a
next-to-leading order (NLO) computation of the dominant matrix elements 
which contribute  to $\widdif$.
\section{Standard Model formula for the width difference in the $\bs$ system.}
The theoretical expression for the width difference reads
\cite{ben97}
\begin{eqnarray}\label{dgg}
\widdif&=&
\frac{G_F^2 m_b^2}{12\pi}f^2_{B_s}M_{B_s}
|V_{cs}V_{cb}|^2\tau_{B_s}\cdot \nonumber\\
&&\hspace{-2cm}\left(G(z)\frac{8}{3} B(m_b)+
    G_S(z)\frac{5}{3}
    B_S(m_b) {\cal X}+\delta_{1/m_b}\right),
\end{eqnarray}
where  
\be\label{Xi}
{\cal X}=\frac{M_{B_s}^2}{(m_b+m_s)^2}\qquad ,
\ee
$z=m_c^2/m_b^2$, $G(z)$ and $G_S(z)$ are NLO Wilson coefficients calculated in
\cite{ben98}, $\delta_{1/m_b}$ are $O(1/m_b)$ corrections which depend on matrix
elements of higher dimension operators (the explicit expression for
$\delta_{1/m_b}$ can be found in \cite{ben97}).\\
$B$ and $B_S$ are bag parameters defined as vacuum insertion deviations,
\begin{eqnarray}\label{qb}
\langle\bar B_s|O_L(\mu)|B_s\rangle &\equiv& \frac{8}{3}f^2_{B_s}M^2_{B_s} B(\mu), \\
\label{qsbs}
\langle\bar B_s|O_S(\mu)|B_s\rangle &\equiv& -\frac{5}{3}f^2_{B_s}M^2_{B_s}{\cal X}\, B_S(\mu) 
\end{eqnarray}
where 
\beqns
O_L&=&\bar b^i\, \gamma^{\mu}\, (1 -\gamma_{5})\, q^i\,  
\, \bar b^j\, \gamma_{\mu}\, (1 -\gamma_{5})\, q^j\\
O_S &=& \bar b^i\, (1-\gamma_{5})\, s^i\, 
 \bar b^j\,  (1-\gamma_{5})\, s^j
 \eeqns
 i, j are color indices.

One of the most important sources of uncertainty in (\ref{dgg}) is the $B_s$
decay constant, $f_{B_s}$. 
In order to reduce it we follow ref. \cite{APE00} and write 
$\widdif$ in the following way,
\begin{eqnarray}\label{dgg2}
\widdif\!\!\!\!\!&=&\!\!\!\!\!
\frac{m_{B_s}}{m_{B_d}}\frac{4\pi}{3}\frac{m_b^2}{m_W^2}
\left|\frac{V_{cb}V_{cs}}{V_{td}V_{tb}}\right|^2
\frac{\tau_{B_s}\,\Delta m_{B_d}}{\eta_B(m_b)\,S_0(x_t)}\xi^2\,
\nonumber\\
&&\hspace{-.5cm}\cdot\left(G(z)
    +G_S(z){\cal R}(m_b)+
   \tilde{\delta}_{1/m}\right),
\end{eqnarray}
Notice that all the non-perturbative contribution, at leading order in $1/m_b$, comes
from the $\cal R$ parameter,
\be
{\cal R}(m_b)\equiv \frac{\langle\bar B_s|{O_S(m_b)}|B_s\rangle}
{\langle\bar B_s|{O_L(m_b)}|B_s\rangle}=
-\frac{5}{8}\frac{B_S(m_b)}{B(m_b)}{\cal X},
\ee
The parameter
$\xi=\frac{f_{B_s}\sqrt{\hat{B}_{B_s}}}{f_{B_d}\sqrt{\hat{B}_{B_d}}}$, 
is accurately computed from lattice simulations \cite{gim&mar96},
$x_t=m_t^2/m_W^2$ and, $\eta_B(m_b)$ and $S_0(x_t)$ are well known theoretically
\cite{buras90,inami81}.
Finally  $\Delta m_{B_d}$ and $\tau_{B_s}$ are accurately measured
experimentally.
\section{NLO matching from continuum QCD to lattice HQET.}
Our aim is to compute the two B-parameters $B$ and $B_S$ from lattice
simulations. The b quark, however, is too heavy to be simulated dynamically
in accessible simulations since the typical inverse lattice spacings, are in the range
$2\div3\; \mathrm{Gev}$. There are two ways to overcome this problem:
performing
simulations around the charm mass and extrapolate the results to the physical b
quark mass or using an effective field theory (HQET, NRQCD) in which the heavy
degrees of freedom have been integrated out. In our work we follow the second
method using HQET. Our task is 
to extract matrix elements of operators in QCD from lattice HQET simulations.
To obtain a NLO result the following four steps, described in detail
in \cite{vicent97}, should be computed:
\begin{description}
\item[1. The continuum QCD--HQET matching.]
The QCD operators are expressed as linear combinations of
HQET ones in the continuum at a given high scale, say, $\mu=m_{b}$.
\item[2. Running down to $\mu=a^{-1}$ in the HQET.]
The HQET operators obtained in step 1 at the high scale $\mu=m_{b}$ are 
evolved down to a lower scale $\mu=a^{-1}$, appropriate for lattice simulations,
using the HQET NLO renormalization group equations, which implies the
computation of the two loop anomalous dimension of the relevant operators. 
\item[3. Continuum--lattice HQET matching.]
Having obtained the continuum HQET operators at the scale $\mu=a^{-1}$,
they are expressed as a linear combination of lattice 
HQET operators at this scale.
\item[\parbox{7cm}{4. Lattice computation of the matrix elements.}]
The matrix elements of the lattice HQET operators in Step 3, are measured by 
Monte Carlo numerical simulations on the lattice. 
\end{description}
We stress that the first three steps depend on the renormalization scheme used
in the HQET. When the two loop anomalous dimensions are included in the
renormalization group evolution (RGE) 
this scheme dependence must cancel up
to $O(\alpha_s)$. So far, the complete chain of matching equations have been
calculated for $O_L$ only. However, the two last steps are known for all operators
\cite{gim&mar96,dipierro98,nos99}. We have extended the calculation to all the
$\Delta B=2$ 
operators computing the continuum QCD -- HQET matching and the two
loop anomalous dimensions \cite{our1,our2}.
We have explicitly verified 
the cancellation of the dependence of our expressions on the intermediate HQET
scheme \cite{our1}. This is a strong check of our calculation of the matching.

\section{B-parameters from numerical simulations on the lattice.}
The lattice B-parameters are extracted from the large time behavior
of the ratio between three- and two-point correlation functions (see for example
\cite{gim&mar96}):
\[
R_{O_i}\equiv\frac{C_{O_i}(-t_1,t_2)}{C(-t_1)C(t_2)}\stackrel{t_1,t_2\to
\infty}{\longrightarrow} 
\frac{\langle\bar B_{Ps}|O_i(a)|B_{Ps}\rangle}{|\langle
0| A_0(a)|B_{Ps}\rangle|^2}
\]
where
\beqns
C_{O_i}(t_1,t_2)\!\!\!\!\!&\equiv&\!\!\!\!\!\!\!\!\sum_{\vec{x}_1,\vec{x}_2} \langle
0|A_0(\vec{x}_1,t_1) O_i(\vec{0},0) A^{\dagger}_0(\vec{x}_2,t_2)|0\rangle\\
C(t)&\equiv&\sum_{\vec{x}} \langle
0| A_0(\vec{x},t) A^{\dagger}_0(\vec{0},0)|0\rangle
\eeqns
and $A^{\mu}$ is the HQET axial current.

The results presented here are obtained from a quenched simulation performed by
APE collaboration in a $24^3\times 40$ lattice with 600 gauge configuration at
$\beta=6.0$ with $a^{-1}=2$ $\mathrm{Gev}$. The details of
the simulation can be found in \cite{gim&mar96} and references therein.

An important remark we want to make is that, as firstly pointed out in
\cite{dipierro98}, all lattice B-parameters obtained in the
static limit are very close to their vacuum insertion values. This result has
been obtained in static simulations by APE coll. \cite{gim&mar96}, UKQCD coll.
\cite{UKQCD}, Christensen et al. \cite{kentucky} and also by JLQCD coll. in
NRQCD when their results are extrapolated to the infinite mass limit
\cite{JLQCD}. Notice that although these simulations are performed with different
actions (static-clover, static-Wilson, NRQCD-clover) at different lattice
spacings and by different groups, the conclusion
is the same in all cases: the lattice B-parameters in the static
limit are compatible with the vacuum insertion value within the statistical
errors. This surprising property of the HQET deserves further investigation
which is in progress.

\section{$\mathbf{1/m_b}$ dependence of the B-parameters.}
Having performed the calculation described in the previous section one obtains
the B-parameters in QCD to leading order in $1/m_b$. 
By using
vacuum insertion approximation (VIA) for the subleading operators
\cite{mannel} and the fact that the bare
lattice 
B-parameters are very close to the VIA value, we have shown \cite{our1} that all
$\Delta B=2$ B-parameters, defined as vacuum 
insertion deviations, have small $O(\bar
\Lambda/m_b)$ corrections ($\bar \Lambda \equiv M_{B_s}-m_b$), 
\be
B_i(m_b)=\bar B_i(m_b) + \underbrace{O(0.3\frac{\bar \Lambda}{m_b})}_{\sim
0.05}+O(\frac{1}{m_b^2}),
\ee
where $B_i(m_b)$ and $\bar B_i(m_b)$ are the full and the static QCD
B-parameters respectively. The factor $0.3$ is an estimation of the deviation
from VIA in the calculation of the subleading operators and
$O(\alpha_s$) corrections.  
\section{Results.}
Before presenting our results, let us stress that $m_s$ and $m_b$ parameters in 
previous sections are the corresponding pole masses. The latter coincides with 
the expansion parameter of the HQET because we have set the  residual mass to
zero. 
We calculate the $b$ quark pole mass from the running $\overline{\mathrm{MS}}$
mass, which can be accurately determined. Since our computation is
performed at NLO, we use the perturbative relation between the pole and the
running mass at the same order. From the world average running
mass \cite{gim00}, $\overline m_b(\overline m_b)=4.23\pm 0.07\,\,\mbox{Gev}$ one
obtains $m_b=4.6\pm 0.1\,\mbox{Gev}$. Notice that the contribution of $m_s$ is
very small because it always appears divided by $m_b$. 

\begin{table}[t]
\parbox{7cm}{\caption{Values of the B-parameters in the
NDR-$\overline{\mathrm{MS}}$ scheme (see text).}\label{tabl1}}\\
\begin{tabular}{cl}
\hline 
\multicolumn{2}{c}{B-parameters}\\
\hline 
\hline 
$B(m_b)$&0.83(5)(4)(5)\\
$B_S(m_b)$&0.96(8)(5)(5)\\
$B^{LR}(m_b)$&0.94(5)(4)(5)\\
$B_S^{LR}(m_b)$&1.03(3)(5)(5)\\
\hline 
\hline 
\end{tabular}
\end{table}

In table \ref{tabl1} we present the
values for all $\Delta B=2$ B-parameters. We have included, for the sake of
completeness, the other $\Delta B=2$ operators in the HQET,
\beqn
\frac{\langle\bar B_s|O^{LR}(\mu)|B_s\rangle}
{ -2f^2_{B_s}M^2_{B_s}}
\left(1+\frac{2}{3}{\cal X}\right)^{-1} &\equiv& B^{LR}(\mu), \\
\frac{\langle\bar B_s|O_S^{LR}(\mu)|B_s\rangle}
{\frac{1}{3}f^2_{B_s}M^2_{B_s}}
\left(1+6\,{\cal X}\right)^{-1} &\equiv&  B_S^{LR}(\mu),
\eeqn
and
\beqn
O^{LR}&=&\bar b^i\, \gamma^{\mu}\, (1 -\gamma_{5})\, q^i\,  
\, \bar b^j\, \gamma_{\mu}\, (1 +\gamma_{5})\, q^j,\\
O_S^{LR}&=& \bar b^i\, (1-\gamma_{5})\, q^i\, 
 \bar b^j\,  (1+\gamma_{5})\, q^j.
 \eeqn
The first error in table 1 comes from lattice
simulations and includes statistical and systematic errors. The second one is an
estimate of the error due to the uncertainties in the values of the lattice
coupling constant and to higher order contributions to the matching. The third
is an estimate of $1/m_b$ corrections to the static result (see eq.(7)).
The 
analysis of the 
perturbative matching is explained in detail in our previous work
\cite{nos99} where we gave
 $B(m_b)=0.81\pm 0.05\pm 0.04$. The tiny difference is due to the fact that
 there we used, in the perturbative  
evolution, a number of flavours $n_f=4$ instead of $n_f=0$ as in the
present paper. The B-parameters in table \ref{tabl1} are QCD scheme dependent,
they all are in the NDR-$\barms$ scheme.
In order to make 
contact with the NLO computation of \cite{ben98}, we have subtracted the
evanescents operators in the renormalization of $O$ and $O_S$ as in \cite{ben98}.
For $B^{LR}$ and  $B_S^{LR}$ we use the prescription of \cite{buras00}.

For the parameter $\cal R$ (see eq.(6)), which contains all the non-perturbative
contribution to the width difference, 
at leading order in $1/m_b$,
we obtain
\be
{\cal R}(m_b)=-0.95(7)(9)
\ee
where as usual the first error comes from lattice simulations, and the second is
systematic due to the uncertainty in the perturbative matching and to
the contribution of higher orders in $1/m_b$.
\begin{table} 
\parbox{7cm}{\caption{Values of  $B$, $B_S$ and ${\cal R}$ from different
groups.}\label{tabl2}}\\ 

\scalebox{.844}{\begin{tabular}{clll}
\hline
\hline
\mbox{\rm Group}&\mbox{$B(m_b)$}&\mbox{$B_S(m_b)$}&\mbox{${\cal R}(m_b)$}
\\
\hline
\cite{JLQCD} &
$ 0.85(3)(11)$&$0.95(2)(12)$&$-0.91(5)(17)$
\nonumber\\\hline
\cite{APE00}
&$ 0.91(3)\err{0.00}{0.06}$&
$ 1.02(2)\err{0.02}{0.04}$&
$ -0.93(3)\err{0.00}{0.06}$\nonumber\\\hline
 This work &$ 0.83(5)(6)$&$ 0.96(8)(7)$&
$ -0.95(7)(9)$\\
\hline
\hline
\end{tabular}}
\end{table}
\subsection{Comparison with other recent results}
In table \ref{tabl2} we present our result for the B-parameters relevant for
$\widdif$ compared with other recent determinations. In \cite{APE00} the $B_S$
is defined as in eq.(\ref{qsbs}) but in terms of
the running mass instead of the pole mass. Therefore, we have multiplied their
value by $(m_b/\overline{m_b}(\overline m_b))^2$ in order to compare with our
result. 
On the other hand, in the definition used in \cite{JLQCD} does not appear the
factor 
${\cal X}$ (see eq. (\ref{Xi})), therefore we have divided their result
 by
${\cal X}$. Notice that in spite of using different methods to obtain the
B-parameters, there is a good agreement between the three computations, in
particular in the value of the ratio ${\cal R}$.

From equation (\ref{dgg2}) we get our prediction,
\be
\widdif=(5.1\pm 1.9\pm 1.7)\times 10^{-2}.
\ee
The first error is systematic obtained from the spread of values of
all input parameters in eq.(\ref{dgg2}). The second one comes from the
uncertainty in the value of $\tilde \delta_{1/m_b}$. Since in the estimate of
$\tilde \delta_{1/m_b}$ the operator matrix elements were computed using VIA
 and the radiative corrections were not included, we assume an error of $30\%$
 \cite{APE00}. As can be seen, this parameter is still affected by a large
 uncertainty, so that a precise determination of the width difference requires
 the computation of the subleading matrix elements using lattice QCD. This
 simulation is missing to date.

Our result is to be compared with the present experimental status \cite{schn}
\be
\left(\frac{\Delta\Gamma_{B_s}}{\Gamma_{B_s}}\right)^{\mbox{exp.}}=\left(17\errd{09}{10}\right)\times
10^{-2}
\ee
As can be seen the central values are rather different but still compatible
within the large errors. More work is needed on the theoretical and experimental
sides to reduce the uncertainty in the width diference in the $B_s$ system.

\end{document}